\documentclass[english,aps,preprint]{revtex4}
\usepackage[]{fontenc}
\usepackage[latin9]{inputenc}
\usepackage{graphicx}
\usepackage{amssymb}
\usepackage{babel}

\begin{document}

\title{Dusty plasma (Yukawa) rings}

\author{T. E. Sheridan}

\email{t-sheridan@onu.edu}

\author{J. C. Gallagher}

\affiliation{Department of Physics and Astronomy, Ohio Northern University, Ada,
Ohio 45810 U.S.A.}

\begin{abstract}
One-dimensional and quasi-one-dimensional strongly-coupled dusty plasma
rings have been created experimentally. Longitudinal (acoustic) and
transverse (optical) dispersion relations for the 1-ring were measured
and found to be in very good agreement with the theory for an unbounded
straight chain of particles interacting through a Yukawa (i.e., screened
Coulomb or Debye-H\"uckel) potential. These rings provide a new system
in which to study one-dimensional and quasi-one-dimensional physics.
\end{abstract}
\maketitle
One-dimensional (1D) chains of particles are an important model system
for studying linear and nonlinear phenomena such as waves, instabilities,
and diffusion \cite{ash,tod,fpu}. A standard model is that of an
unbounded straight chain of particles interacting through a given
force. This system may represent a lumped model of a continuous system,
such as a string, or may model the atomic lattice. Although infinite
systems cannot be realized in practice, they can be approximated using
large systems with periodic boundary conditions so that end effects
are eliminated. One experimental system where it is possible to study
strongly-coupled 1D chains is dusty (complex) plasma.

Dusty plasma is a system of microscopic material particles immersed
in an electron-ion plasma. For typical laboratory conditions, the
dust particles acquire a net charge $q<0$ and can be confined in
a single-layer suspension above a horizontal electrode. Vertical confinement
is due to the balance between the downward gravitational force and
the average upward electrostatic force inside the plasma sheath. Horizontal
confinement is often due to depressions in the plasma potential created
by secondary electrodes. The interaction between dust particles is
screened by the response of the electrons and ions so that the potential
a distance $r$ from a dust particle is given by a Yukawa (screened
Coulomb or Debye-H\"uckel) potential \cite{lam,kon}\begin{equation}
V\left(r\right)=\frac{1}{4\pi\epsilon_{0}}\,\frac{q}{r}e^{-r/\lambda_{D}},\label{eq:Yuk}\end{equation}
where $\lambda_{D}$ is the Debye screening length. In principle,
the interaction length can be varied from long to short range allowing
study of both {}``plasma'' and {}``condensed matter'' regimes
\cite{tes7}. Since the dominant friction force on dust particles
is neutral gas (Epstein) drag, wave modes may be underdamped at low
gas pressures.

Previous experimental work on 1D dusty plasmas has focused on straight
chains confined in highly anisotropic two-dimensional (2D) biharmonic
wells \cite{hom,liu1,liu2,melz,mis,tes2}. Such systems have several
drawbacks. First, the lattice constant $a$ is not a constant and
increases at the ends of the chain. Second, the boundary conditions
are not periodic, so that comparisons to unbounded theories are problematic.
Third, it is difficult to make very long straight chains since 1D
chains are unstable against the zigzag when the lattice constant is
below a critical value \cite{melz,tes2}. To overcome these drawbacks
it may be possible to create a one-dimensional Coulomb ring \cite{schw}
or dusty plasma ring \cite{tes1} using a two-dimensional (2D) annular
potential well. A dusty plasma ring will provide the most direct test
of the theory for lattice waves in unbounded Yukawa chains \cite{mela},
as well as a system with which to explore phenomena such as single-file
diffusion \cite{nel} and the properties of quasi-one-dimensional
systems \cite{pia}.

In the present work we demonstrate that it is possible to create one-dimensional
and quasi-one-dimensional dusty plasma rings in experiment, thereby
providing a real system that models a chain of interacting particles
with periodic boundary conditions. We measure the longitudinal and
transverse dispersion relations for a 1-ring using thermal fluctuations,
and confirm that the waves are well described by the theory of an
unbounded 1D Yukawa chain.

We briefly review dusty plasma ring theory \cite{tes1}. The dusty
plasma consists of $n$ identical particles with mass $m$ and charge
$q$ at positions $\left\{ x_{i},y_{i}\right\} $ confined in a 2D
annular potential well \cite{schw}. The potential energy of an isolated
particle in this well is\begin{equation}
U_{{\rm well}}\left(r\right)=\frac{1}{2}m\omega_{0}^{2}\left(r-s\right)^{2},\label{eq:Ur}\end{equation}
where $\omega_{0}$ is the single-particle radial oscillation frequency,
$r$ is a radial coordinate and $s$ is the radius at which the well
potential energy is a minimum. The total potential energy of a particle
configuration is\begin{equation}
U=\frac{1}{2}m\omega_{0}^{2}\sum_{i=1}^{n}\left(\sqrt{x_{i}^{2}+y_{i}^{2}}-s\right)^{2}+\frac{q^{2}}{4\pi\epsilon_{0}}\sum_{i<j}^{n}\frac{e^{-r_{ij}/\lambda_{D}}}{r_{ij}},\label{eq:U}\end{equation}
where the second sum is the interaction potential energy for all unique
pairs of particles and $r_{ij}$ is the distance between particles
$i$ and $j$. The potential energy $U$ is nondimensionalized by
defining variables\begin{equation}
\kappa=\frac{r_{0}}{\lambda_{D}},\;\sigma=\frac{s}{r_{0}},\;\xi_{i}=\frac{x_{i}}{r_{0}},\;\eta_{i}=\frac{y_{i}}{r_{0}},\;\rho_{ij}=\frac{r_{ij}}{r_{0}},\label{eq:ndvars}\end{equation}
where the characteristic length and energy are\begin{equation}
r_{0}^{3}=\frac{1}{\frac{1}{2}m\omega_{0}^{2}}\,\frac{q^{2}}{4\pi\epsilon_{0}},\; U_{0}^{3}=\frac{1}{2}m\omega_{0}^{2}\left(\frac{q^{2}}{4\pi\epsilon_{0}}\right)^{2},\label{eq:r0U0}\end{equation}
to give\begin{equation}
\frac{U}{U_{0}}=\sum_{i=1}^{n}\left(\sqrt{\xi_{i}^{2}+\eta_{i}^{2}}-\sigma\right)^{2}+\sum_{i<j}^{n}\frac{e^{-\kappa\rho_{ij}}}{\rho_{ij}}.\label{eq:ndU}\end{equation}
In dimensionless form, time-independent configurations $\left\{ \xi_{i},\eta_{i}\right\} $
are determined by three parameters: $n$, $\kappa$ and $\sigma$.
Here $\kappa$ is the Debye shielding parameter and $\sigma$ is the
radius at which the confining well has its minimum. This model was
shown \cite{tes1} to have one-dimensional ring solutions (i.e., 1-rings).
A 1-ring is unstable against the zigzag instability when the particle
density exceeds a critical value, which may lead to the creation of
quasi-1D systems consisting of concentric rings, e.g., 2-rings.

For a 1D chain with lattice parameter $a$ the dimensionless lattice
parameter is $\alpha=a/r_{0}$. The curvature of the ring is negligible
when $s\gg a$ and $s\gg\lambda_{D}$. In this case, an appropriate
shielding parameter is\begin{equation}
\bar{\kappa}=\frac{a}{\lambda_{D}}=\alpha\kappa.\label{eq:kbar}\end{equation}
The dispersion relation for an undamped longitudinal (acoustic) wave
in an unbounded straight 1D Yukawa chain is \cite{mela} \begin{equation}
\frac{\omega_{l}^{2}}{\omega_{0}^{2}}=\frac{4}{\alpha^{3}}\sum_{j=1}^{2}\frac{e^{-j\bar{\kappa}}}{j^{3}}\left[1+j\bar{\kappa}+\frac{1}{2}\left(j\bar{\kappa}\right)^{2}\right]\sin^{2}\left(\frac{jka}{2}\right),\label{eq:wl}\end{equation}
where $\omega_{l}$ is the longitudinal wave frequency and $k$ is
the real wavenumber. The dispersion for the transverse (optical) mode
is \cite{liu1}\begin{equation}
\frac{\omega_{t}^{2}}{\omega_{0}^{2}}=1-\frac{1}{\alpha^{3}}\sum_{j=1}^{\infty}\frac{e^{-j\bar{\kappa}}}{j^{3}}\left(1+j\bar{\kappa}\right)\left[1-\cos\left(jka\right)\right],\label{eq:wt}\end{equation}
where $\omega_{t}$ is the transverse wave frequency. The optical
mode is a backward wave. These dispersion relations were previously
found to be in good agreement with theoretical normal mode spectra
for a dusty plasma 1-ring \cite{tes1}.

\begin{figure}
\includegraphics[width=3in]{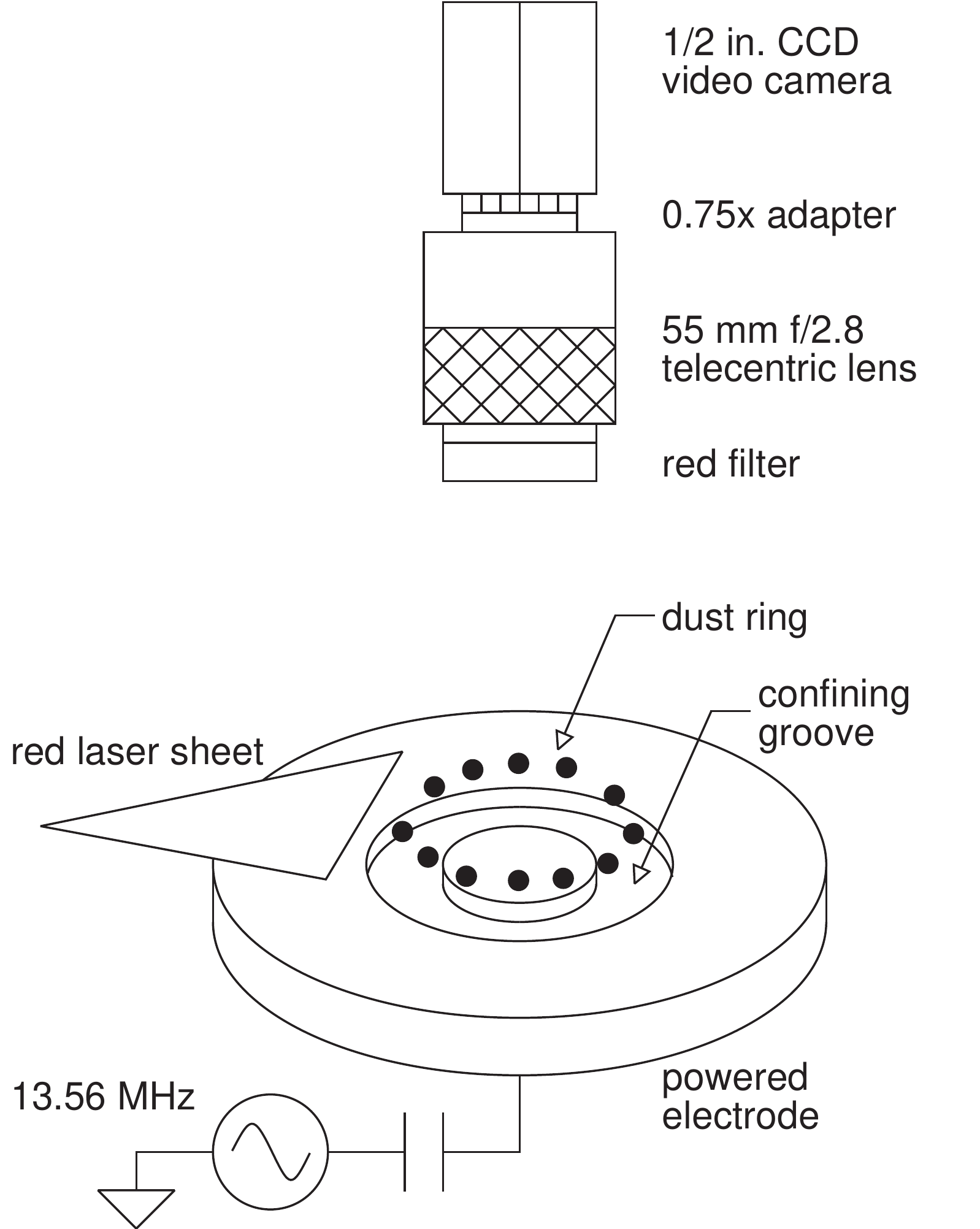}

\caption{\label{fig:scheme}Schematic of the experimental setup. An annular
groove with rectangular cross section cut into the rf electrode produces
an annular confining well for dust particles. When the particle spacing
is above a critical value the particle configuration is one-dimensional
with a ring topology. }

\end{figure}

Our experimental setup is shown in Fig. \ref{fig:scheme}. Monodisperse
microscopic ($8.94\pm0.09\;\mu{\rm m}$ diameter) melamine dust particles
are levitated in an rf plasma discharge above the capacitively-coupled
powered electrode in the DONUT (Dusty ONU experimenT) apparatus \cite{tes2,tes3,tes4,tes5,tes6}.
Dust confinement is due to a 2D annular potential well created by
cutting an annular rectangular groove in the surface of the 89-mm
powered electrode. The groove has an inner diameter of 15.9 mm, an
outer diameter of 40.0 mm and a depth of 3.3 mm. The plasma sheath
edge partially conforms to the groove, creating an annular depression
in the equipotential surfaces above the groove. The dust particles
are illuminated using a steady sheet of 635-nm laser light. Particle
positions are recorded using a video camera. For these experiments
1025 frames of 1024 $\times$ 1024 pixels were recorded at 7.51 frames/s. 

\begin{figure}
\includegraphics[width=3.25in]{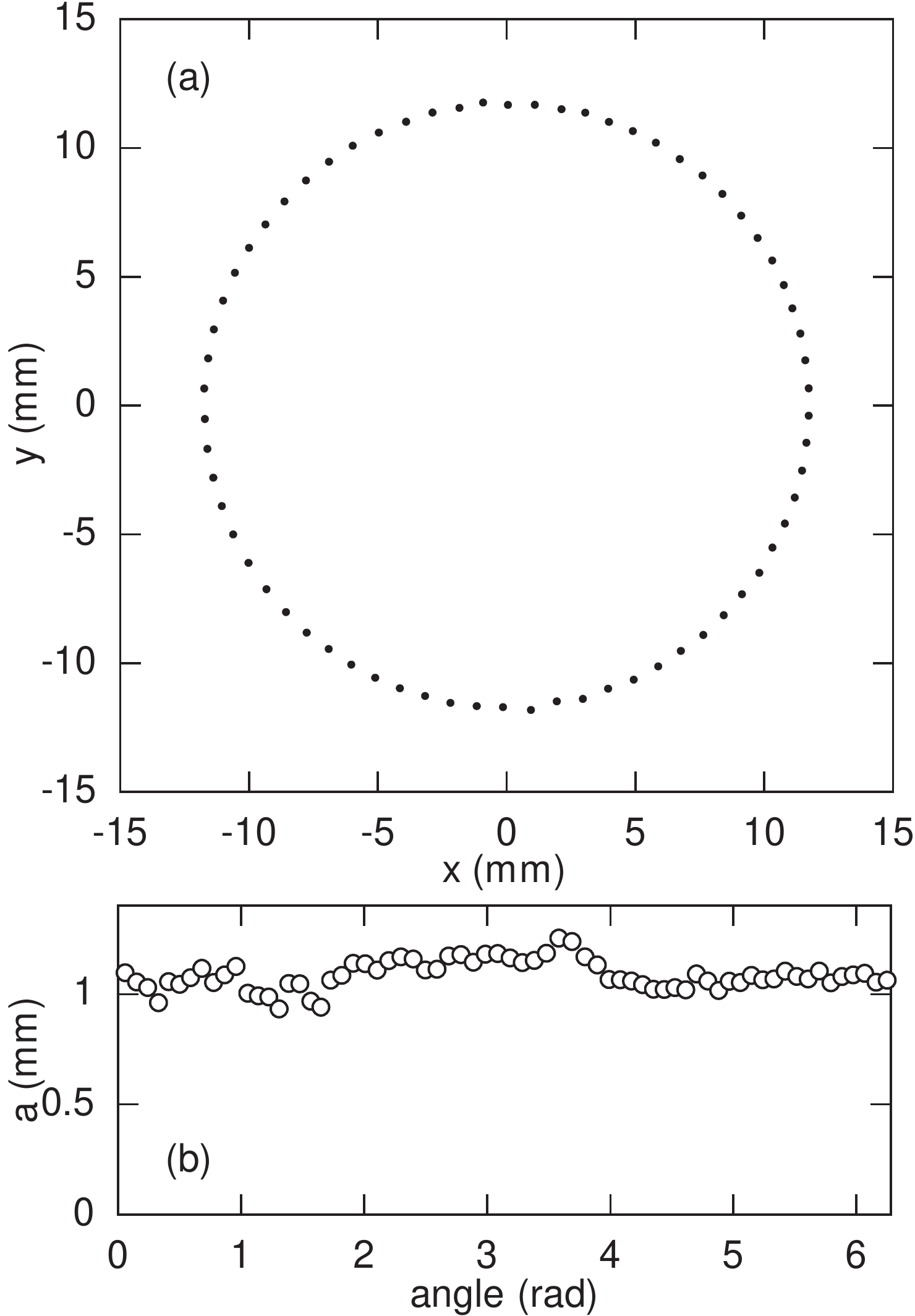}

\caption{\label{fig:ring}(a) Measured equilibrium positions of $n=68$ particles
confined in a two-dimensional annular well. The particles form a one-dimensional
ring. (b) Distance $a$ between nearest neighbors vs azimuthal angle.}

\end{figure}

Making a uniform dusty plasma ring {[}Fig. \ref{fig:ring}] requires
the bottom of the potential well to be flat. This condition is more
stringent than confinement in a parabolic well since a tilted parabolic
well is still parabolic. To create a dust ring we adjusted the tilt
of the electrode (actually the entire vacuum vessel) until the particles
were distributed approximately uniformly in a ring. However, at low
pressures ($\lesssim20$ mtorr) we observed an $m=2$ azimuthal perturbation
to the potential well. The exact characteristics of this perturbation
depend on the rf power, suggesting it may be due to an rf standing-wave
component. Such a perturbation cannot be eliminated by tilting the
electrode. We used two grounded L-shaped rods inserted horizontally
above the electrode to cancel this perturbation by varying the distance
and orientation of the rods. 

An equilibrium configuration of a dusty plasma 1-ring is shown in
Fig. \ref{fig:ring}(a). The ring consists of $n=68$ dust particles
floating 9.2 mm above the top of the electrode. Discharge parameters
were a neutral pressure of 18.9 mtorr argon, $\sim3$ W forward rf
power at 13.56 MHz, and the dc self-bias on the electrode was -52
V. Wave modes are underdamped at this pressure. The camera resolution
was $0.0317\;{\rm mm/pixel}$ . To determine the geometric center
of the ring we fitted the particle coordinates to a circle \cite{sha}
and then averaged over all frames. (Because of variable particle spacing
the geometrical center and the center of mass are not identical.)
We found an average radius $R=11.73\;{\rm mm}$, so that the average
particle separation computed as the circumference over $n$ is $a=1.084\;{\rm mm}$.
The dust ring lies closer to the inner edge of the groove. The Euclidean
distance $r_{ij}$ between adjacent pairs of particles is shown in
Fig. \ref{fig:ring}(b). The average separation $\left\langle r_{ij}\right\rangle =1.084\pm0.008$
mm agrees with the previous estimate of $a$.

\begin{figure}
\includegraphics[width=3.25in]{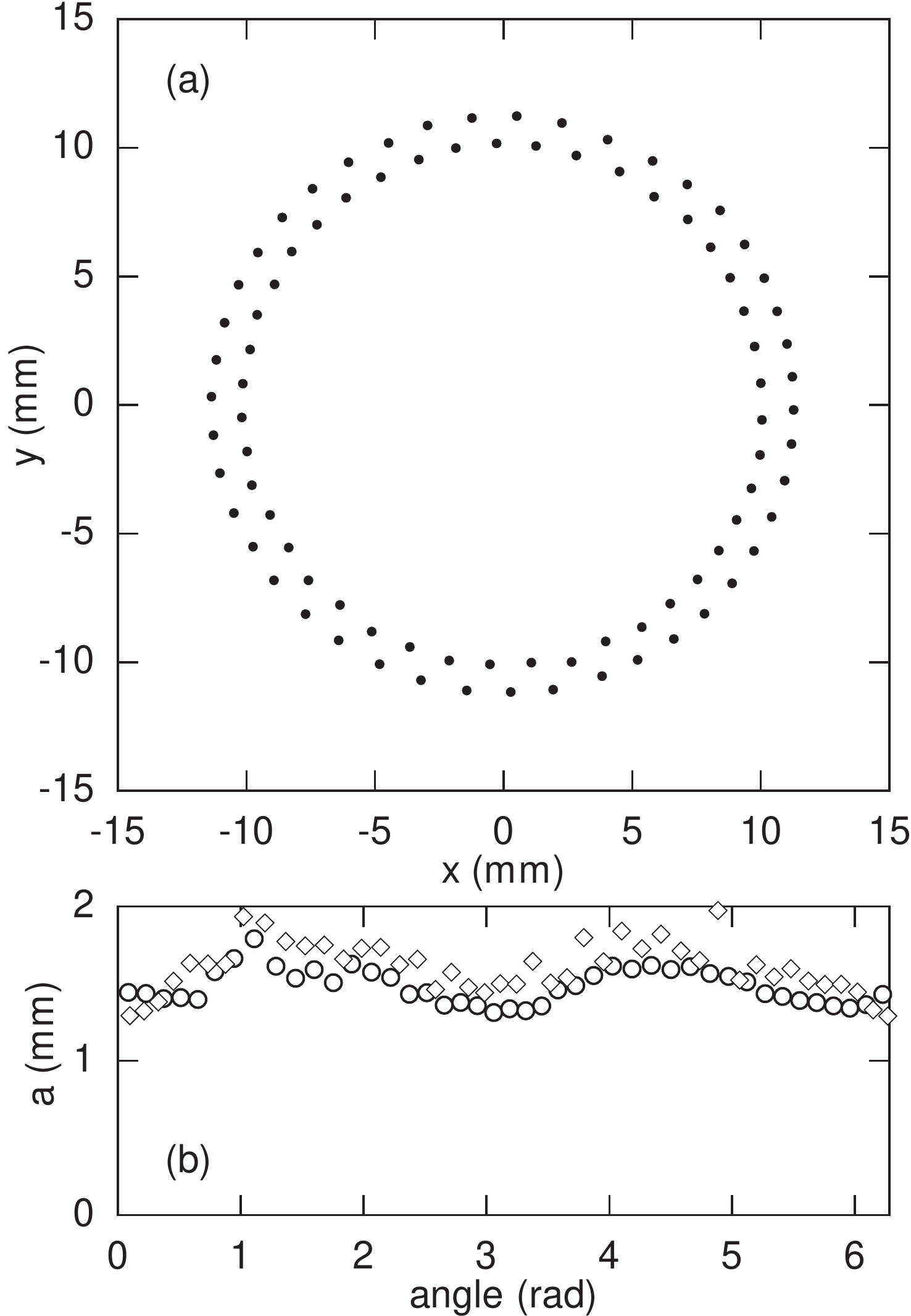}

\caption{\label{fig:2ring}(a) Quasi-one-dimensional configuration (a 2-ring)
consisting of $n=87$ particles arranged in two concentric rings with
43 particles in the inner ring and 44 in the outer ring. (b) Distance
between particles in the inner ring (open circles) and in the outer
ring (open diamonds) vs angle.}

\end{figure}

In Fig. \ref{fig:2ring}(a) we demonstrate that it is also possible
using this confinement geometry to create a quasi-1D configuration
of two rings (i.e., a 2-ring) with different radii, as predicted by
theory \cite{tes1,schw}. In this case, we have a total of $n=87$
particles, with 43 particles in the inner ring and 44 particles in
the outer ring. The radius of the inner ring is $10.12\pm0.01$ mm,
and that of the outer ring is $11.24\pm0.01$ mm. For most of this
two-ring the particles are in a zigzag configuration. However, near
$0^{\circ}$ the two rings {}``slip'' past each other. The distance
between pairs of particles in the inner and outer rings is plotted
in Fig. \ref{fig:2ring}(b). The average distance between particles
in the inner ring is $1.48\pm0.02$ mm, and $1.60\pm0.03$ mm for
the outer ring. An $m=2$ modulation in the particle spacing is visible.

The particle positions undergo small amplitude thermal fluctuations,
mostly due to their interaction with the neutral gas component \cite{tes3}.
For the 1-ring {[}Fig. \ref{fig:ring}(a)], velocity distributions
in the longitudinal (azimuthal) and transverse (radial) directions
were computed from the time histories of the measured particle positions
\cite{nun}. The velocity distribution functions indicate a minimum
measurable velocity of $\approx0.06$ mm/s, where a velocity of 1
pixel/frame corresponds to $v=0.238\;{\rm mm/s}$. Except for a small
neighborhood around $v=0$, the distributions in both directions are
well fitted by a Maxwellian with a temperature $T=410\;{\rm K}$,
somewhat below 530 K reported for a 2D lattice \cite{nun}. This temperature
indicates that the ring is stable and in equilibrium with the neutral
gas.

The dependence of the power spectral density for the thermal fluctuations
vs wavenumber $k$ for the 1-ring were computed for the longitudinal
(acoustic) and transverse (optical) modes. First, the average position
of each particle was computed by averaging over all 1025 frames. Then,
for each frame the longitudinal and transverse displacements were
computed and Fourier analyzed using a discrete Fourier transform (DFT)
to give the amplitude and phase of each $k$ vs time, which assumes
the equilibrium lattice spacing is independent of position. The power
spectral density (psd) of the time history of each wavenumber was
then computed using the maximum entropy method (MEM) \cite{numrec}.
The psd's from four data sets were averaged to give the final psd.
Since each normal mode acts as an independent damped harmonic oscillator,
the psd peaks near the mode's natural frequency and so the maximum
of the psd should follow the dispersion relations \cite{liu2}.

\begin{figure}
\includegraphics[width=3.5in]{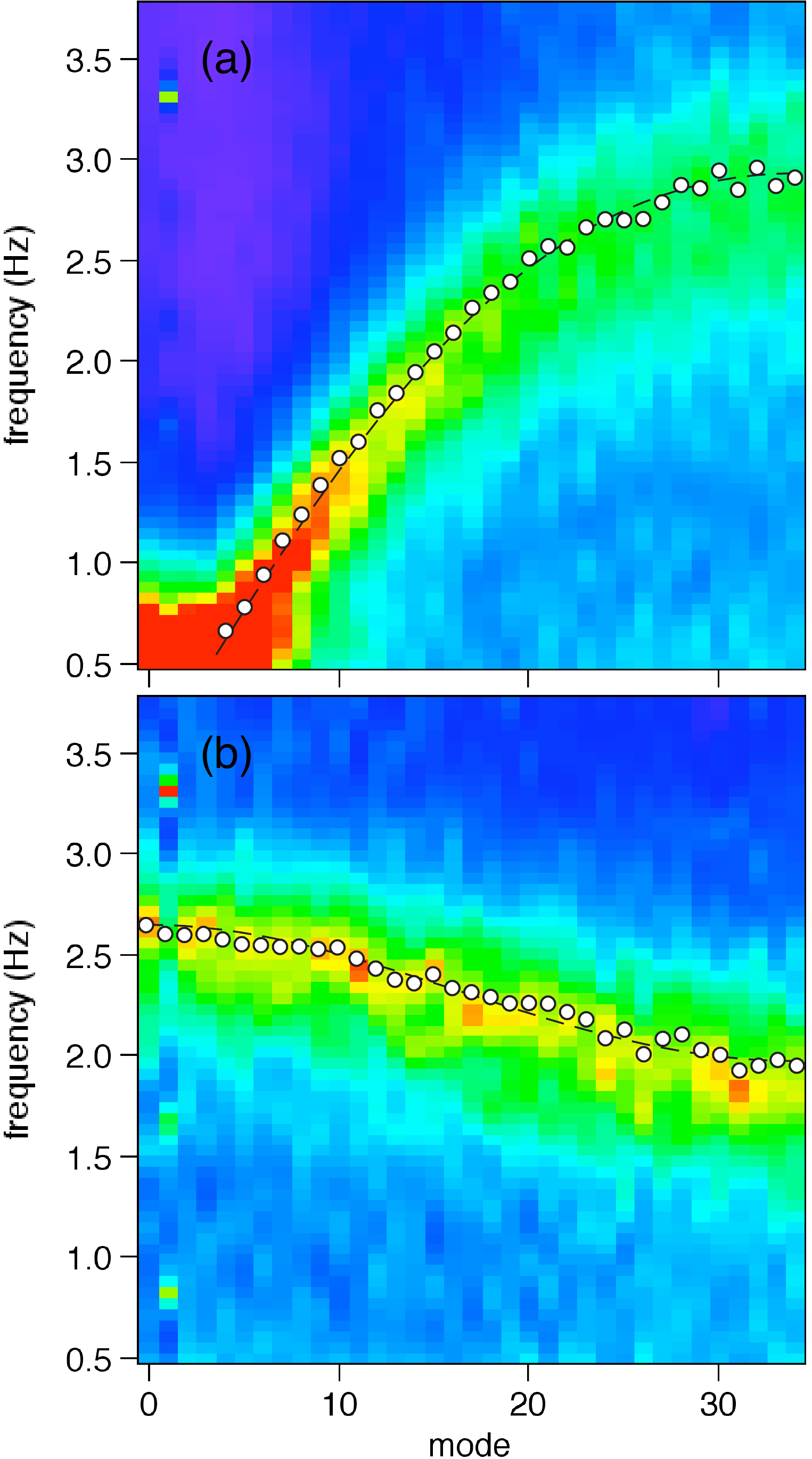}

\caption{\label{fig:psd}(Color online) (a) Power spectral density (psd) for
thermally-excited longitudinal modes vs mode number and frequency
and (b) psd for transverse modes. Purple corresponds to the least,
and red to the greatest, power density. Open circles are the natural
frequency for each mode, and the dashed line is the fit to the theoretical
dispersion relations.}

\end{figure}

The power spectral density for the acoustic modes is displayed in
Fig. \ref{fig:psd}(a). Here the horizontal axis runs from mode number
0 to 34, or equivalently $ka=0$ to $\pi$. The vertical axis starts
at 0.5 Hz because of low-frequency noise and runs to the Nyquist frequency
of 3.75 Hz. The power spectral density for the transverse mode is
shown in Fig. \ref{fig:psd}(b). Here we see a typical optical mode,
with a finite frequency for $k=0$, which should be $\omega_{0}$
{[}Eq. (\ref{eq:Ur})], and then a decrease in $\omega$ with $k$
indicating a backwards wave. The $m=1$ mode displays several narrow
band features due to minute mechanical vibrations coupled to the vacuum
chamber from our belt-drive mechanical pump. In this case, the electrode
and camera oscillate while the dust ring remains stationary.

We fitted the psd for each mode with the equation for a driven damped
harmonic oscillator for both the longitudinal and transverse directions
to determine the natural mode frequency $\omega_{i}$ vs $k$. Here
$\omega_{i}$ may be slightly above the maximum of the psd because
of finite damping, where the damping rate $\gamma\sim2$ s$^{-1}$.
We then fitted both measured dispersion relations simultaneously to
theory {[}Eqs. (\ref{eq:wl}) and (\ref{eq:wt})], where there are
only three free parameters: $\omega_{0}$, $\alpha$ and $\bar{\kappa}$.
From this procedure, we found $\omega_{0}=16.6$ rad/s, $\alpha=1.41$
and $\bar{\kappa}=1.32\pm0.05$. We believe that this is the first
report of very good agreement of experimental data simultaneously
with \emph{both }branches of the dispersion relation, confirming that
this system is well described by a model for an unbounded 1D chain
of particles interacting through a Yukawa potential. For $\bar{\kappa}=1.32$
the critical value of the lattice parameter $\alpha=a/r_{0}$ for
the zigzag instability is $\alpha_{c}=1.13$. For the ring studied
here, $\alpha=1.41$ is significantly above $\alpha_{c}$. Dimensional
parameters are $r_{0}=0.77$ mm, $\lambda_{D}=0.82$ mm and $q=-1.24\times10^{4}e$.
The values of $\lambda_{D}$ and $q$ are consistent with those found
in DONUT for similar plasma conditions \cite{tes2,tes3,tes4,tes5,tes6}. 

In summary, we have created 1D and quasi-1D dusty plasma rings experimentally,
verifying previous predictions for charged particles confined in a
2D annular well \cite{schw,tes1}. Measured longitudinal and transverse
dispersion relations in the 1-ring exhibit excellent agreement with
dispersion relations in an unbounded Yukawa lattice, confirming that
these systems are well described by a model of $n$ identical charged
particles confined in an annular potential and interacting through
a Yukawa potential. These rings provide a new experimental system
for studying static and dynamic properties of 1D and quasi-1D systems.

\end{document}